\documentclass[pra,aps,twocolumn]{revtex4-1}

\usepackage{amsmath} 
\usepackage{amsthm} 
\usepackage{amssymb}	
\usepackage{graphicx} 
\usepackage{subfigure}

\begin{document}
\title{$\mathcal{PT}$-symmetry breaking with divergent potentials: lattice and continuum cases} 
\author{Yogesh N. Joglekar$^1$, Derek D. Scott$^1$, and Avadh Saxena$^2$}
\affiliation{$^1$ Department of Physics, 
Indiana University Purdue University Indianapolis (IUPUI), 
Indianapolis, Indiana 46202, USA}
\affiliation{$^2$ Theoretical Division and Center for Nonlinear Studies, Los Alamos National Laboratory, Los Alamos, New Mexico 87544, USA}

\begin{abstract}
We investigate the parity- and time-reversal ($\mathcal{PT}$)-symmetry breaking in lattice models in the presence of long-ranged, non-hermitian, $\mathcal{PT}$-symmetric potentials that remain finite or become divergent in the continuum limit. By scaling analysis of the fragile $\mathcal{PT}$ threshold for an open finite lattice, we show that continuum loss-gain potentials $V_\alpha(x)\propto i |x|^\alpha \mathrm{sign}(x)$ have a positive $\mathcal{PT}$-breaking threshold for $\alpha>-2$, and a zero threshold for $\alpha\leq -2$. When $\alpha<0$ localized states with complex (conjugate) energies in the continuum energy-band occur at higher loss-gain strengths. We investigate the signatures of $\mathcal{PT}$-symmetry breaking in coupled waveguides, and show that the emergence of localized states dramatically shortens the relevant time-scale in the $\mathcal{PT}$-symmetry broken region. 
\end{abstract}
\maketitle


\section{Introduction}
\label{sec:intro}
Since Bender and co-workers' seminal work on non-hermitian Hamiltonians a decade and a half ago, there has been tremendous progress in the field of parity and time-reversal ($\mathcal{PT}$-) symmetric quantum theory~\cite{bender1,bender2,bender3}. For continuum, $\mathcal{PT}$-symmetric, non-hermitian  Hamiltonians on an infinite line, they showed that the eigenvalue spectrum is purely real when the strength of the ``non-hermiticity'' is small, and becomes complex when it is large. Traditionally, the region of the parameter space where the eigenvalues of the $\mathcal{PT}$-symmetric Hamiltonian are purely real, $\epsilon_\lambda=\epsilon^*_\lambda$, and the eigenfunctions are simultaneous eigenfunctions of the combined $\mathcal{PT}$-operation, $f_\lambda(x)=f^*_\lambda(-x)$, is called the $\mathcal{PT}$-symmetric region. In the early years, significant theoretical progress was made towards the development of a self-consistent quantum theory via a Hamiltonian-dependent inner product, under which the eigenfunctions become orthonormal in the $\mathcal{PT}$-symmetric phase~\cite{bender2}. This progress was accompanied by mathematical advances in the field of pseudohermitian operators - operators that are not hermitian under the standard inner product, but may be self-adjoint under an appropriately defined metric~\cite{mostafa}. Most of these investigations were focused on continuum Hamiltonians on an infinite line. 

During the past five years, discrete $\mathcal{PT}$-symmetric Hamiltonians on finite lattices and continuum $\mathcal{PT}$-symmetric Hamiltonians on a finite line have been extensively studied due to their experimental relevance. $\mathcal{PT}$ symmetry breaking is a {\it non-perturbative phenomenon} that occurs when the strength of the non-hermitian potential is equal to the relevant hermitian energy scale. A perturbation-theory characterization of the $\mathcal{PT}$-symmetry breaking criterion was developed in Ref.~\cite{moiseyev}, which showed that the eigenvalues of a $\mathcal{PT}$-symmetric Hamiltonian $H=H_0 +i\gamma V$ remain real when the strength $\gamma$ of the non-hermitian potential $V$ is smaller than the radius of convergence $\gamma_R$ for the perturbation expansion; this radius is determined by the hermitian piece $H_0$. It also demonstrated that coupled optical waveguides with balanced loss and gain provide an ideal candidate to visualize the effects of approaching the $\mathcal{PT}$-breaking transition~\cite{moiseyev}. Since then, it has become clear that $\mathcal{PT}$-symmetric  Hamiltonians naturally arise as ``effective Hamiltonians'' for open systems with balanced loss and gain, and $\mathcal{PT}$-symmetry breaking experimentally manifests as a transition from a quasiequilibrium state to a state with broken reciprocity. Experimental demonstrations of $\mathcal{PT}$-symmetry breaking in optics~\cite{expt1,expt2,expt3,expt4}, and the natural emergence of $\mathcal{PT}$-symmetric effective potentials in driven condensed matter systems~\cite{ma,microbridge,kamanev} have complemented theoretical studies of $\mathcal{PT}$-symmetry breaking in lattice models~\cite{znojil1,znojil2,bendix,song,avadh,longhi1} and continuum models on a finite line~\cite{znojilpt,makrispt,longhipt,graefept,gunther,wkb1,wkb2}. Special attention has been paid to the number of eigenvalues that become complex~\cite{jake}, their location in the energy spectrum~\cite{mark,charles}, the extended or localized nature of the corresponding eigenstates~\cite{optlett}, and the experimental consequences of the spatial extent of the states that break the $\mathcal{PT}$ symmetry~\cite{defectprl}. 

A salient difference between the lattice and continuum models is as follows. In all cases, continuum models on a finite line have shown a positive $\mathcal{PT}$-symmetry breaking threshold~\cite{znojilpt,makrispt,longhipt,graefept,gunther,wkb1,wkb2}. In contrast, most lattice models have shown a vanishing $\mathcal{PT}$-symmetry breaking threshold that goes to zero as the number of lattice sites $N$ diverges~\cite{bendix,mark,charles,dima}. This remarkable discrepancy suggests that understanding the differences between lattice and continuum models is crucial for a detailed understanding of the $\mathcal{PT}$-symmetry breaking phenomenon, particularly because all of its realizations have been in small lattices with $N\lesssim 100$ sites. 

Here, we investigate $\mathcal{PT}$-symmetry breaking in $N$-site lattices with extended loss-gain potentials characterized by strength $\gamma>0$ and parameter $\alpha$, and their continuum counterparts on a finite segment. The paper is organized as follows. In the next section, we present the tight-binding model, and discuss the results for the $\mathcal{PT}$-symmetric threshold $\gamma_{PT}(N,\alpha)$ on the lattice and their continuum implications. In particular, we show that some divergent continuum potentials on a finite segment have a positive $\mathcal{PT}$-symmetry breaking threshold. In Sec.~\ref{sec:psi} we discuss the signatures of $\mathcal{PT}$-symmetry breaking in such lattices, and show that they are consistent with the expectations based on the extended nature of $\mathcal{PT}$-broken eigenstates. In Sec.~\ref{sec:loc} we show that localized states with complex energies within the lattice energy band emerge at much larger loss-gain strength $\gamma_c\gg \gamma_{PT}$, and discuss their significance. We conclude the paper with Sec.~\ref{sec:disc}. 


\section{Tight-binding model}
\label{sec:tb}
\begin{figure*}[htpb]
\centering
\includegraphics[width=0.49\textwidth]{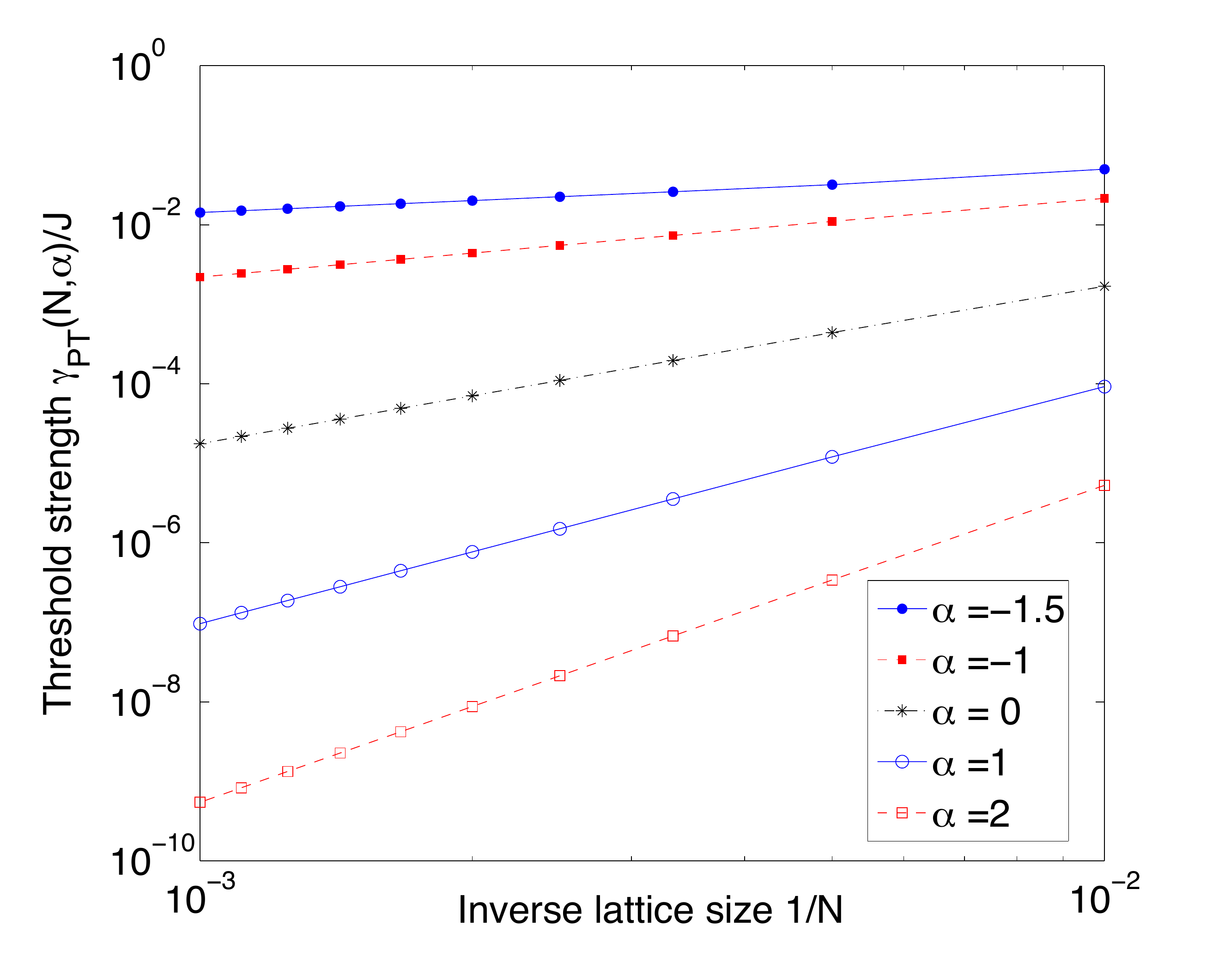}
\includegraphics[width=0.48\textwidth]{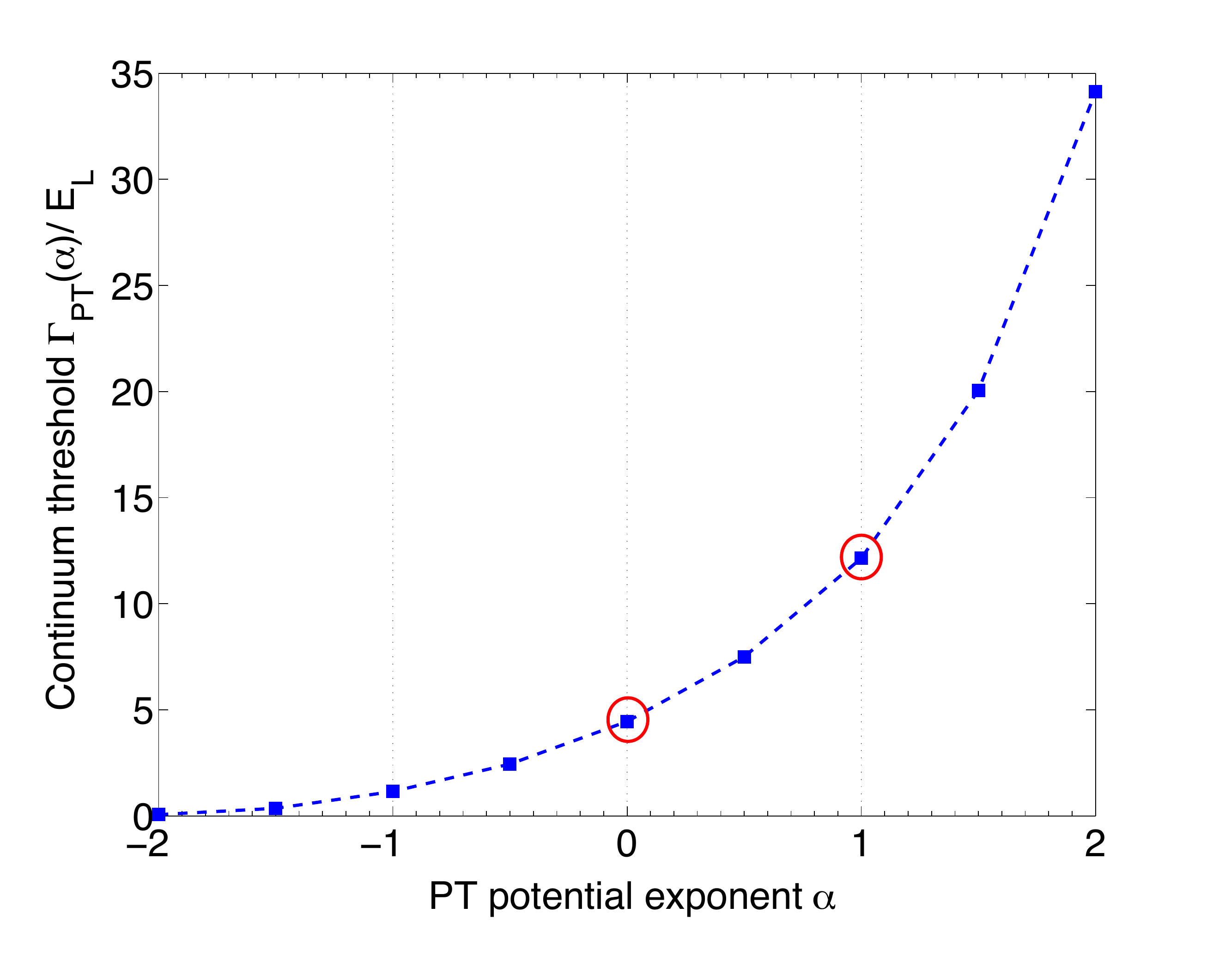}
\caption{(Color online) Left-hand panel: the lattice $\mathcal{PT}$-threshold $\gamma_{PT}(N,\alpha)/J$ shows a power-law dependence on the inverse lattice size $1/N$ for different loss-gain potentials $V_\alpha$. These results are obtained with $100\leq N\leq 1000$. Note the logarithmic scale on both axes. Right-hand panel: the continuum threshold $\Gamma_{PT}(\alpha)/E_L$, obtained from the scaling data, shows that continuum models, Eq.(\ref{eq:calpha}), have a positive threshold for $\alpha>-2$. This includes divergent potentials such as $V(x)=i\Gamma L/x$.}
\label{fig:ptphase}
\end{figure*}
Consider an $N$-site tight-binding lattice with site-to-site distance $a$ and nearest-neighbor tunneling $J>0$. Its hermitian tunneling Hamiltonian is given by 
\begin{equation}
\label{eq:tunneling}
H_{0}=-J \sum_{k=1}^{N-1} \left(  a^\dagger_{k+1} a_k + a^\dagger_k a_{k+1}  \right),
\end{equation}
where $a^{\dagger}_{k}(a^{}_{k})$ represents the creation (annihilation) operator for a state $ |k\rangle$ localized at site $k$. We keep coupled optical waveguides in mind for an experimental realization of this lattice; thus, $a^\dagger_k$ represents the creation operator for the single-mode electric field in the waveguide $1\leq k\leq N$. The parity operator on an open lattice is given by $\mathcal{P}: a^\dagger_n\rightarrow a^\dagger_{\bar{n}}$ where site $\bar{n}=N+1-n$ is the parity-symmetric counterpart of site $n$. The action of the time-reversal (or motion-reversal) operator is $\mathcal{T}:i\rightarrow -i$. We note that the Hamiltonian Eq.(\ref{eq:tunneling}) represents a lattice with open boundaries, and thus its eigenfunctions $\psi_n(m)$ satisfy the constraint $\psi_n(m=0)=0=\psi_n(m=N+1)$. In the continuum limit, this boundary condition translates into Dirichlet boundary condition with a vanishing wave function.

The spectrum of the tight-binding model is given by $E_n=-2J\cos(k_n)$, and the corresponding extended, normalized eigenfunctions consistent with the open boundary condition are $\psi_n(m)=\langle m|\psi_n\rangle=\sin(k_n m)$. Here $k_n=n\pi/(N+1)$ with $1\leq n\leq N$. Note that the spectrum is symmetric about zero, $E_n=-E_{\bar{n}}$, the eigenfunctions have equal weights on parity-symmetric sites, and the eigenstates at energies $\pm E_n$ are related by $\langle m|\psi_n\rangle=(-1)^m \langle m|\psi_{\bar{n}}\rangle$. These symmetries of the spectrum and eigenfunctions remain valid in the presence of pure loss-gain potentials in the $\mathcal{PT}$-symmetric region~\cite{ph}. 

We consider a class of extended loss-gain potentials parameterized by $\alpha$, 
\begin{equation}
\label{eq:alpha}
V_{\alpha}=i \gamma\sum_{k=1}^N |k-n_c|^{\alpha}\mathrm{sign}(k-n_c)a^{\dagger}_k a_k.
\end{equation}
Here, $\gamma>0$ is the strength of the potential and $n_c=(N+1)/2$ is the lattice center. With the  present convention, the first half of the lattice, $k\leq n_c$, is the ``loss region'' and the second half of the lattice, $k>n_c$, is the ``gain region''. In coupled optical waveguides, such a potential is implemented by a site-dependent complex index of refraction $n_k=n_{Rk}+i n_{Ik}$ with a symmetric real part, $n_{Rk}=n_{R\bar{k}}$, and an antisymmetric imaginary part, $n_{Ik}=-n_{I\bar{k}}$. At this point, we remind the reader that extended potentials on a lattice, Eq.(\ref{eq:alpha}), were investigated and deemed unstable due to the vanishing $\mathcal{PT}$-symmetric threshold that is obtained in the limit $N\gg 1$~\cite{bendix,dima}.

The continuum limit of this problem is defined by $N, J\rightarrow\infty$ and $a\rightarrow 0$ in such a manner that $Na\rightarrow 2L$ defines the length of the finite segment and $Ja^2\rightarrow\hbar^2/2m$ defines the mass of the non-relativistic quantum particle confined in this segment.  With this notation, it follows that the continuum potential becomes  
\begin{equation}
\label{eq:calpha}
V_\alpha(x) =  i\Gamma\mathrm{sign}(x)\left|\frac{x}{L}\right|^\alpha=V_\alpha^*(-x), 
\end{equation}
where the potential strength $\Gamma$ is given by 
\begin{equation}
\label{eq:Gamma}
\Gamma = \lim_{N\rightarrow\infty} \gamma\left(\frac{N}{2}\right)^\alpha.
\end{equation}
Note that the continuum potential Eq.(\ref{eq:calpha}) is not analytic at $x=0$ except when $\alpha$ is an odd integer. Then, for $\alpha>0$, $V_\alpha(x)$ reduces to cases considered in earlier investigations~\cite{bender1,bender2,bender3,gunther,wkb1,wkb2}, although none of those works consider divergent potentials $\alpha<0$. The continuum problem corresponding to the Hamiltonian $H_0+V_\alpha$ is given by the Schr\"{o}dinger equation 
\begin{equation}
\label{eq:schr}
-\frac{\hbar^2}{2m}\partial_x^2 \psi_q(x) + V_\alpha(x) \psi_q(x) = \mathcal{E}_q \psi_q(x)
\end{equation}
subject to boundary conditions $\psi_q(x=\pm L)=0$. At this point, we remind the reader that a positive $\mathcal{PT}$-breaking threshold was found for the spectrum of Eq.(\ref{eq:schr}) when $\alpha=0$~\cite{znojilpt}, $\alpha=1$~\cite{ma,microbridge,gunther},  and $\alpha=3, 5$~\cite{wkb1,wkb2}. In this section, we will resolve this apparent discrepancy between the lattice and continuum results. 

Since the spectrum of the $\mathcal{PT}$-symmetric Hamiltonian $H_\alpha=H_0+V_\alpha$ cannot be obtained analytically, we numerically obtain the threshold $\gamma_{PT}(N,\alpha)$ below which all eigenvalues of the discrete $N\times N$ Hamiltonian are purely real. The left-hand panel in Fig.~\ref{fig:ptphase} shows the dependence of the threshold on the lattice size. We see that $\gamma_{PT}(N,\alpha)/J$ decreases in a power-law fashion as $N$ increases, and that the power-law exponent is determined by $\alpha>-2$. Thus, for $\alpha>-2$, the $\mathcal{PT}$-symmetric threshold vanishes with increasing $N$ and the resultant $\mathcal{PT}$-symmetric phase is fragile~\cite{dima}, but in a very specific manner,  
\begin{equation}
\label{eq:trend}
\frac{\gamma_{PT}(N,\alpha)}{J}\rightarrow A_\alpha\left(\frac{2}{N} \right)^{\alpha+2}, 
\end{equation}
for $N\gg 1$. It follows from Eqs.(\ref{eq:Gamma}) and (\ref{eq:trend}) that the dimensionless continuum threshold for $V_\alpha(x)$ is equal to the power-law prefactor,
\begin{equation}
\label{eq:cont}
\frac{\Gamma_{PT}(\alpha)}{E_L}=\lim_{N\rightarrow\infty} \frac{\gamma_{PT}}{J} \left(\frac{N}{2}\right)^{\alpha+2}=A_\alpha, 
\end{equation}
where $E_L=\hbar^2/2mL^2$ is the continuum energy scale for a particle on a finite segment. This continuum threshold $\Gamma_{PT}(\alpha)/E_L$, obtained from the scaling data, is shown in the right-hand panel of Fig.~\ref{fig:ptphase}. It shows that $\Gamma_{PT}(\alpha)/E_L$ is a positive, monotonically increasing function of $\alpha$ that goes to zero as $\alpha\rightarrow -2^{+}$. These results imply that, surprisingly, {\it divergent potentials including $V(x)=i\Gamma L/x$ have a positive, finite $\mathcal{PT}$- breaking threshold.} 

When $\alpha\leq -2$, we find that the numerically obtained lattice threshold $\gamma_{PT}(\alpha)/J$ is independent of the lattice size $N$. Therefore, the corresponding continuum threshold obtained via Eq.(\ref{eq:cont}) vanishes, $\Gamma_{PT}(\alpha\leq-2)/E_L=0$. 

To quantify the applicability of the scaling proposed in Eq.(\ref{eq:trend}) to finite lattices, in Table~\ref{table:alpha} we list the mean value of $A_\alpha$ and its variance obtained from the size-dependent $\gamma_{PT}(N,\alpha)$ for lattice sizes varying in steps of 200 from $N=200$ to $N=2000$. We see that for $\alpha\geq -1$, the variance in $A_\alpha$ is less than 1\% of its mean value; for $\alpha=-1.5$, the variance is larger, but only due to finite-size effects that become dominant as the exponent $\alpha+2\rightarrow 0$ when $\alpha\rightarrow -2$. Thus, the scaling trend postulated in Eq.(\ref{eq:trend}) holds well down to $N\sim$ few hundred. The second column in Table~\ref{table:alpha} also indicates that when $\alpha=\{0,1,3,5\}$, the mean $A_\alpha$ exactly matches the continuum threshold results obtained in the literature. 
\begin{table}[h]
\begin{tabular}{|c|c|c|c|}
\hline
\hspace{2mm} exponent $\alpha$\hspace{2mm} & \hspace{2mm} mean $A_\alpha$ \hspace{2mm} & \hspace{2mm} $A_\alpha$ variance \hspace{2mm} & \hspace{3mm} ratio \hspace{3mm}
\\ \hline
-1.5 & 0.3259 & 0.0109 & 0.033 \\ \hline
-1.0 & 1.1092 & 0.0079 & 0.007 \\ \hline
-0.5 & 2.4046 & 0.0118 & 0.005 \\ \hline
0.0	& 4.4436~\cite{znojilpt} & 0.0231	& 0.005 \\ \hline
0.5	& 7.5454	& 0.0253 & 0.003 \\ \hline
1.0	& 12.2470~\cite{ma} & 0.0510 & 0.004 \\ \hline
1.5	& 20.2661 & 0.0866 & 0.004 \\ \hline
2.0	& 34.4561 & 0.2430 & 0.007 \\ \hline
2.5 	& 40.9759 & 0.2454 & 0.006 \\ \hline
3.0	& 50.9557~\cite{wkb1} & 0.3314 & 0.007 \\ \hline
3.5	& 69.0344 & 0.4891 & 0.007 \\ \hline
4.0	& 89.8517 & 0.7237 & 0.008 \\ \hline 
4.5	& 102.1642 & 0.8520 & 0.008 \\ \hline
5.0	& 121.6964~\cite{wkb2} & 1.1069 & 0.009\\ \hline
\end{tabular}
\caption{The mean and variance of $A_\alpha$ obtained from vanishing lattice thresholds $\gamma_{PT}(N,\alpha)/J$ with $200\leq N\leq 2000$.}
\label{table:alpha}
\end{table}

Results in Fig.~\ref{fig:ptphase} and Table~\ref{table:alpha} reconcile the nonzero $\mathcal{PT}$-threshold, i.e. $\Gamma_{PT}(\alpha)/E_L>0$, in a continuum model with a vanishing $\mathcal{PT}$-threshold,  i.e. $\gamma_{PT}(\alpha)/J\rightarrow 0$, in the corresponding lattice model. The existence of a positive continuum threshold, particularly for $\alpha\geq 0$, follows from perturbation theory. The spectrum of the hermitian Hamiltonian $H_0$ is given by $\epsilon_n=E_L (n\pi/2)^2$ and has a finite minimum gap. $\mathcal{PT}$-symmetry breaking is preceded by the closing of the {\it finite gap between adjacent eigenvalues}, which, in turn requires a finite strength of the non-hermitian potential~\cite{langer}. This finite threshold mandates, via Eq.(\ref{eq:Gamma}), that the corresponding lattice-model threshold $\gamma_{PT}/J$ must vanish algebraically with increasing lattice size. These findings are not surprising for $\alpha\geq 0$, when the continuum potential $V_\alpha(x)$ is bounded over the entire line. However, our analysis also predicts that divergent $\mathcal{PT}$-potentials, too, have a positive threshold when $-2<\alpha<0$. (Such potentials have not been investigated in the literature.) In the following section, we investigate the signatures of $\mathcal{PT}$ symmetry breaking in such potentials. 


\section{PT breaking signatures}
\label{sec:psi}

\begin{figure*}
\centering
\includegraphics[width=0.49\textwidth]{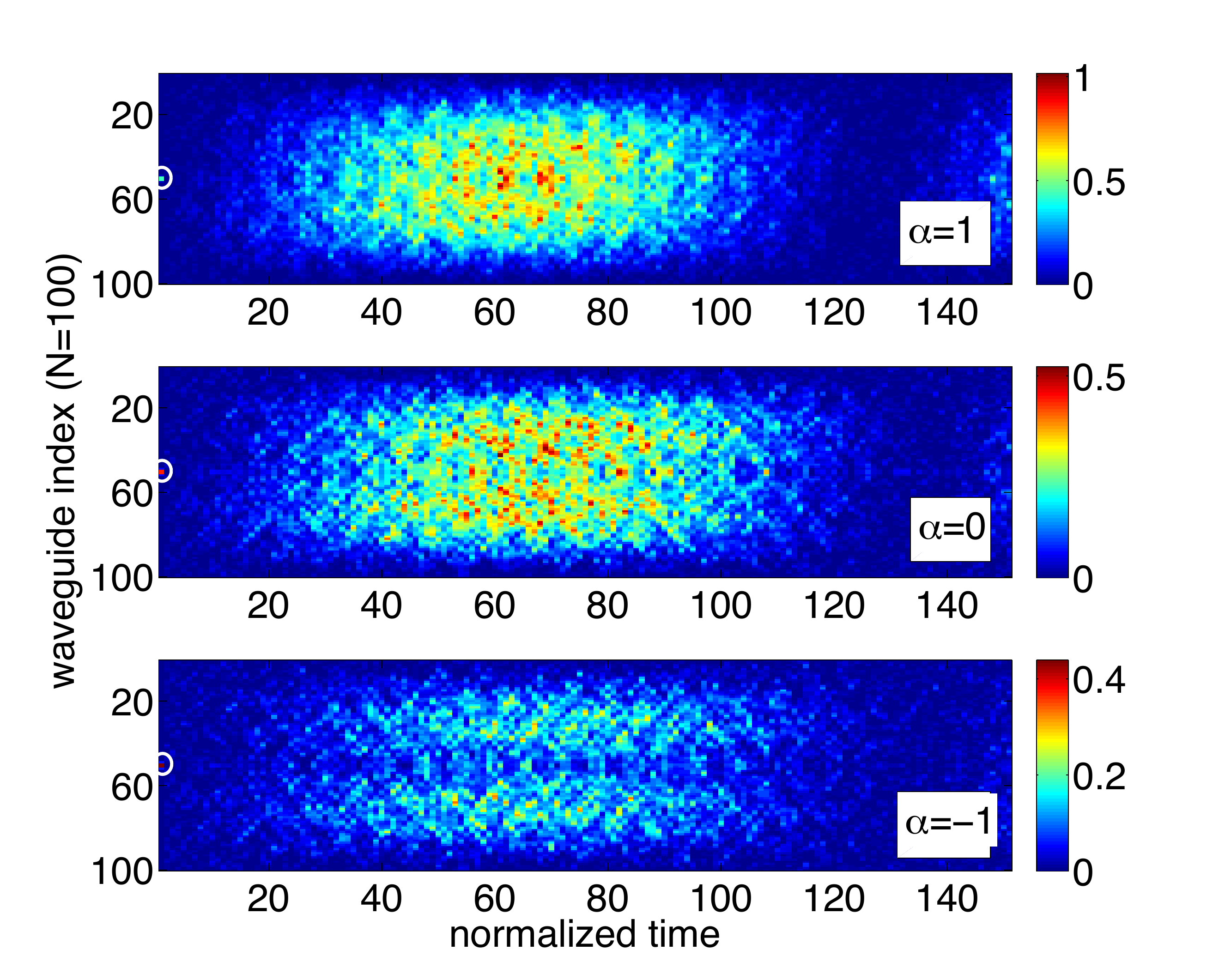}
\includegraphics[width=0.49\textwidth]{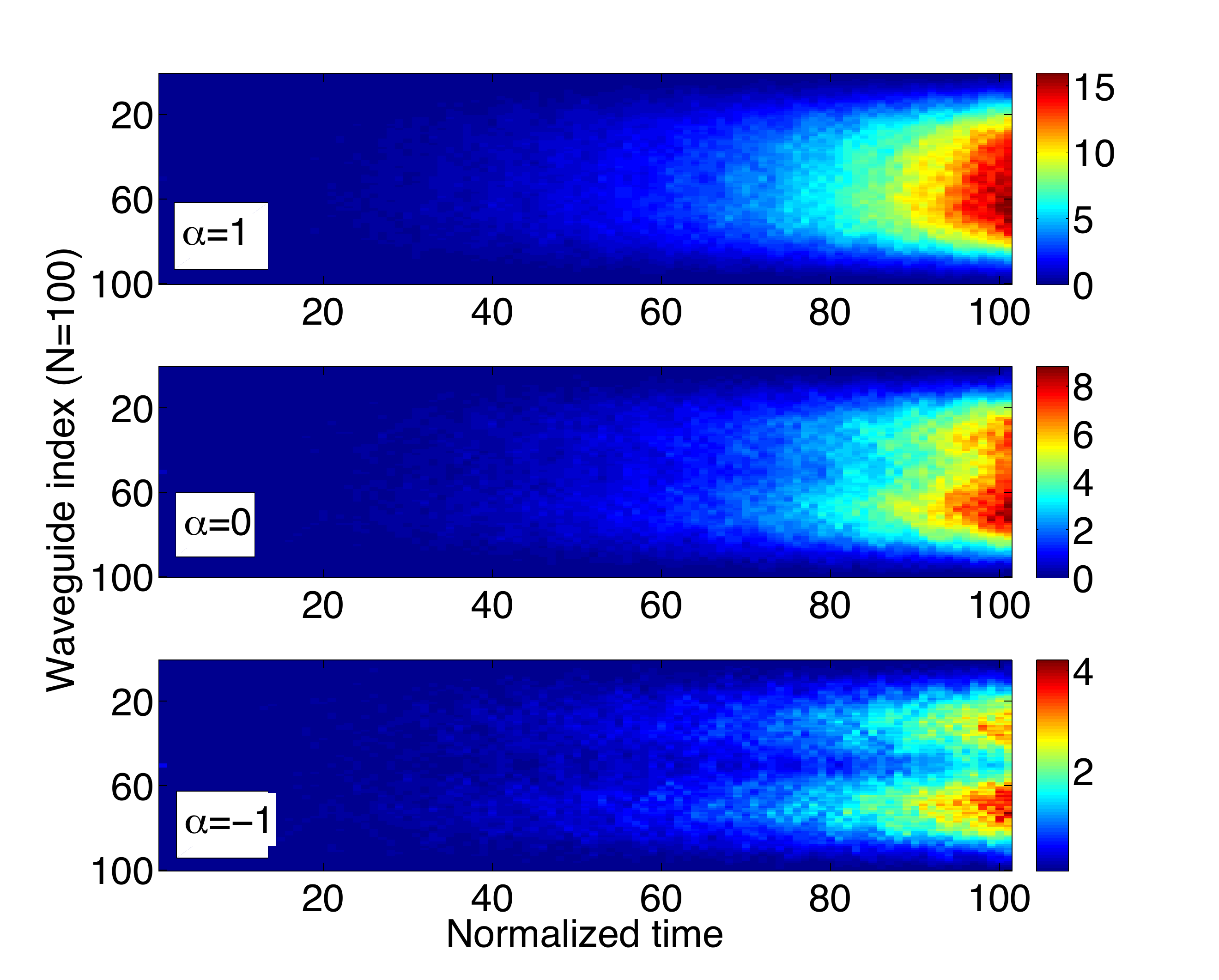}
\caption{(Color online) Signatures of $\mathcal{PT}$-breaking transition in extended potentials with $\alpha=1$ (top row), $\alpha=0$ (center row), and $\alpha=-1$ (bottom row). In each panel, the time is normalized in units of inverse average level spacing $T_s=2\pi/\Delta_{\mathrm{av}}$. Left-hand column: below the threshold, $\gamma/\gamma_{PT}(N,\alpha)=0.995$, the wave packet undergoes bounded intensity oscillation, its intensity in the loss region lags the intensity in the gain region, and the maximum intensity {\it does not occur} in the region with largest loss-gain potential. Right-hand column: after the threshold, $\gamma/\gamma_{PT}(N,\alpha)=1.005$, the wave packet intensity increases monotonically, with the loss-region lagging behind the gain-region. The initial state is localized near the center of 
the $N=100$ lattice, Eq.(\ref{eq:psi0}) with $\sigma=1$ and $k_0=n_c$.} 
\label{fig:psitime}
\end{figure*}

The lattice Hamiltonian $H_\alpha$ can be realized in an array of coupled optical waveguides. The tunneling $J$ is determined by the waveguide cross-section and the distance between adjacent waveguides. It is easily tuned with present-day technology~\cite{review1,reviewlonghi,review2}, as is the loss potential in the first half of the lattice, $k\leq n_c$, engineered via the imaginary part of the index of refraction. The fabrication of an extended, position-dependent gain potential has not yet been experimentally demonstrated, although it may be relatively straightforward to implement in the discrete parity-time synthetic lattices~\cite{expt4,defectprl}. In this section, we present the signatures of $\mathcal{PT}$-symmetry breaking in the time-evolution of an initially normalized wave packet, and discuss their relationship with the spatial structure of the $\mathcal{PT}$-broken eigenfunctions. 

Since $H_\alpha$ is a single-particle, time-independent Hamiltonian, it is straightforward to obtain the time-evolved state $|\psi(t)\rangle=G(t)|\psi(0)\rangle$ where $G(t)=\exp(-iH_\alpha t/\hbar)$ is the non-unitary time evolution operator. The site-and time-dependent intensity is then obtained as $I(k,t)=|\langle k|\psi(t)\rangle|^2$, and the total intensity $I(t)=\sum_{k} I(k,t)$ is not conserved. Note that since the finite-dimensional Hamiltonian $H_\alpha(\gamma)$ is diagonalizable for $\gamma\neq\gamma_{PT}$, the net intensity $I(t)$ oscillates but remains bounded when $\gamma<\gamma_{PT}$ and increases exponentially with time at long times when $\gamma>\gamma_{PT}$. At the $\mathcal{PT}$-breaking point, the Hamiltonian is defective, and can be reduced to a Jordan canonical form with at least one non-trivial Jordan block. Therefore, at long times, the net intensity scales as a power-law, $I(t)\propto t^{2(p-1)}$, where $p\geq 2$ is the dimension of the nontrivial Jordan block corresponding to the $\mathcal{PT}$-breaking, degenerate eigenvalue; at shorter times, the intensity $I(t)$ is a polynomial of order $2(p-1)$ whose exact form is determined by the Hamiltonian at the exceptional point $\gamma=\gamma_{PT}$. Thus, for any finite lattice, at the $\mathcal{PT}$-breaking threshold, {\it the net intensity at long times scales as an even power of time} or, equivalently, the distance along the waveguide. We will see in the next section that the relevant time-scale that codifies ``long time'' is crucially determined by the extended vs. localized nature of eigenfunctions with complex energies. 

We use an initial state centered at lattice site $k_0$,
\begin{equation}
\label{eq:psi0}
\langle k|\psi(0)\rangle= \frac{1}{A}e^{-(k-k_0)^2/2\sigma^2},
\end{equation}
where $A^2(\sigma,N,k_0)=\sum_{k=1}^N \exp\left[-(k-k_0)^2/\sigma^2\right]$ ensures that the state is normalized. The results shown in Fig.~\ref{fig:psitime} are for $\sigma=1$ and $k_0=n_c$, but one obtains {\it qualitatively similar results} for broad wave packets centered at arbitrary locations. The left-hand column in Fig.~\ref{fig:psitime} shows the site- and time-dependent intensity for $\alpha=+1$ (top panel), $\alpha=0$ (center panel), and $\alpha=-1$ (bottom panel). In each case, the $\mathcal{PT}$-potential strength is just below the threshold, $\gamma/\gamma_{PT}(N,\alpha)=0.995$, the unit of time is given by  $T_s=2\pi/\Delta_{\mathrm{av}}$ where the average level spacing is $\Delta_{\mathrm{av}}\sim 4J/N$ ($\hbar=1$). We have chosen the time-range $t/T_s\leq 150$ to cover one bounded intensity oscillation.  The top panel shows that when $\alpha=+1$, the wave packet undergoes amplification near the center of the lattice. The region of maximal intensity is spread out broadly for $\alpha=0$ (center panel), whereas when $\alpha=-1$, the maximum amplification does not occur near the center of the lattice. In each case maximum site-intensity in the loss region, $k\leq n_c$, lags the maximum site-intensity in the gain region, $k>n_c$. The right-hand column shows that just above the threshold, $\gamma/\gamma_{PT}(N,\alpha)=1.005$, the intensity profile monotonically increases with time, but retains an identical $\alpha$-dependence.  

These intensity profiles have two surprising features. The first is that the amplification in the gain region is faithfully transferred to the loss region. Thus, the lag between the intensity maxima is the primary distinguisher between gain and loss regions. The second is that the maximum site intensity {\it does not occur in the region of maximum gain potential.} When $\alpha>0$, the loss-gain potential maxima occur at the two ends of the lattice and the intensity maxima from the non-unitary time evolution are concentrated near the center. When $\alpha<0$, the loss-gain maxima occur at the center of the lattice, whereas the intensity maxima are displaced outward. This counterintuitive behavior is understood by focusing on the eigenfunctions that break the $\mathcal{PT}$-symmetry and dominate the non-unitary time evolution. At $\gamma=0$, the ground state wave function $\psi_1(k)=\sin(\pi k/(N+1))$, and the excited state wave function $\psi_2(k)=\sin(2\pi k/(N+1))$ are orthogonal. As $\gamma\rightarrow\gamma_{PT}$, the corresponding $\gamma$-dependent eigenfunctions become degenerate, and show a maximum at the center of the lattice for $\alpha>0$ and a dip at the center of the lattice for $\alpha<0$. Indeed, at long times, the intensity profile $I(k,t)$ is determined by the intensity profile of these eigenfunctions and is therefore (mostly) independent of the choice of the initial state $|\psi(0)\rangle$. 

For all potentials $V_\alpha$ considered in this paper, the $\mathcal{PT}$-symmetry is first broken via the ground-state and first-excited-state eigenfunctions that are extended over the entire lattice. When the potential strength exceeds the fragile threshold $\gamma_{PT}(N,\alpha)\propto J (2/N)^{\alpha+2}$ and continues to increase, generically, the fraction of eigenvalues that become degenerate and then complex increases, and the corresponding extended eigenfunctions become $\mathcal{PT}$ asymmetric~\cite{gunther,langer}. In the following section, we will investigate the emergence of {\it localized states with complex energies} at loss-gain potential strengths comparable to the lattice bandwidth,  $\gamma\sim J\gg \gamma_{PT}(N,\alpha)$. 

\section{Bound states in the continuum}
\label{sec:loc}
The discrete spectrum of the Hamiltonian $H_0$ is bounded by $\pm 2J$ and becomes a continuous band when the lattice is infinite. A remarkable property of the Hamiltonian $H_\alpha$ is that, no matter how ``strong'' the loss-gain potential is, the real part of the energy spectrum of $H_\alpha$ remains confined to this band while the imaginary part of complex energies increases with the loss-gain strength for $\gamma\gg\gamma_{PT}$. Just as bound states occur in the presence of a hermitian potential, they do in the presence of a non-hermitian, $\mathcal{PT}$-symmetric potential when its strength $\gamma$ exceeds a threshold $\gamma_c$. The crucial difference in the latter case, though, is that the (real part of) energy of such localized states lies in the band $\pm 2J$. It is possible to analytically obtain the threshold $\gamma_c$ for a single pair of $\mathcal{PT}$ impurities in an infinite lattice~\cite{optlett}. However, in the present case with extended potentials $V_{\alpha}$, we locate this threshold numerically. 
\begin{figure}[htpb]
\centering
\includegraphics[width=\columnwidth]{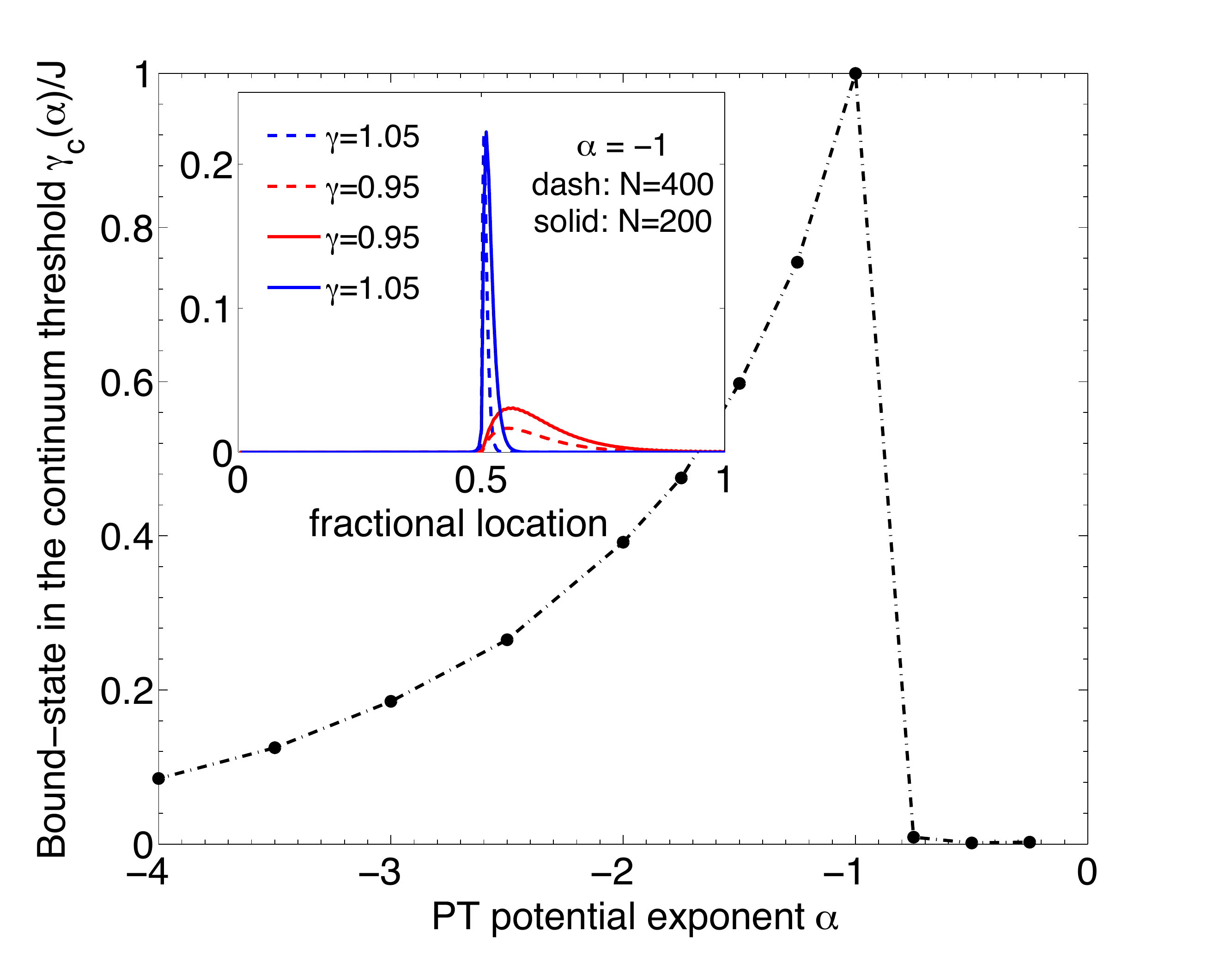}
\caption{(Color online) Bound-state threshold $\gamma_c(\alpha)/J$ for an infinite lattice, obtained from lattice sizes $N$=100-4000. There are no localized states for a positive $\alpha$. The threshold is vanishingly small for $-1<\alpha\leq 0$, and when $\alpha<-1$ the threshold decreases monotonically with $\alpha$. The inset shows the emergence of a localized eigenstate for $\alpha=-1$ with $\gamma_c=J$. An extended eigenstate at $\gamma/\gamma_c=0.95$ (red solid or dashed line) becomes localized past the threshold, $\gamma/\gamma_c=1.05$ (blue solid or dashed line). The vertical axis in the inset is the eigenstate site intensity.}
\label{fig:loc}
\end{figure}

Figure~\ref{fig:loc} shows the bound-state threshold $\gamma_c(\alpha)/J$ for an infinite lattice,  obtained from the data for lattice sizes ranging from $N=100$ to $N=4000$. When $\alpha>0$, the potential does not have any bound states. As $\alpha$ decreases, the potential deepens near the center and bound states localized near the lattice-center emerge. We find that the requisite threshold $\gamma_c(\alpha)/J$ decreases monotonically with decreasing $\alpha\leq -1$. This is consistent with the fact that for large, negative $\alpha$, the potential is concentrated near the lattice center, and therefore the threshold strength necessary to support a bound state is lowered. When $-1<\alpha<0$, the cumulative gain 
\begin{equation}
\label{eq:cumul}
V_c(\alpha)=\gamma\sum_{k>n_c} ^N \frac{1}{(k-n_c)^{|\alpha|}} 
\end{equation}
diverges as $N\rightarrow\infty$ for any value of $\gamma>0$. Therefore the threshold strength $\gamma_c(\alpha)$ required for a bound state vanishes in this limit. We emphasize that the bound-state threshold value is much larger than the $\mathcal{PT}$-symmetry breaking threshold for the same lattice, $\gamma_c(\alpha)/J\sim 1\gg \gamma_{PT}(\alpha,N)/J$. 

The inset in Fig.~\ref{fig:loc} shows the emergence of a bound state at $\alpha=-1$, when the bound-state threshold is $\gamma_c/J=1$. The horizontal axis represents fractional location along the lattice, and the vertical axis denotes the site intensity; the solid lines represent results for $N=200$ and the dashed lines correspond to the $N=400$ case. When $\gamma/\gamma_c=0.95$, the eigenstate site-intensity is nonzero over the entire gain region. (The eigenstate with the complex-conjugate energy has site-intensity that is nonzero over the entire loss region.) As is expected for an extended state, when the lattice size is doubled from $N=200$ (red solid line) to $N=400$ (red dashed line) the site intensity is reduced by a factor of two. This changes dramatically as $\gamma$ crosses the bound-state threshold. At $\gamma/\gamma_c=1.05$, the corresponding eigenstate now becomes localized near the lattice center in the gain region. We note that in contrast to the extended state, the localized state site-intensity profiles for $N=200$ (blue solid line) and $N=400$ (blue dashed line) have the same height, and the fractional width of the profile is halved as $N$ is doubled. This is the key signature of a localized state. 

\begin{figure}[htpb]
\centering
\includegraphics[width=\columnwidth]{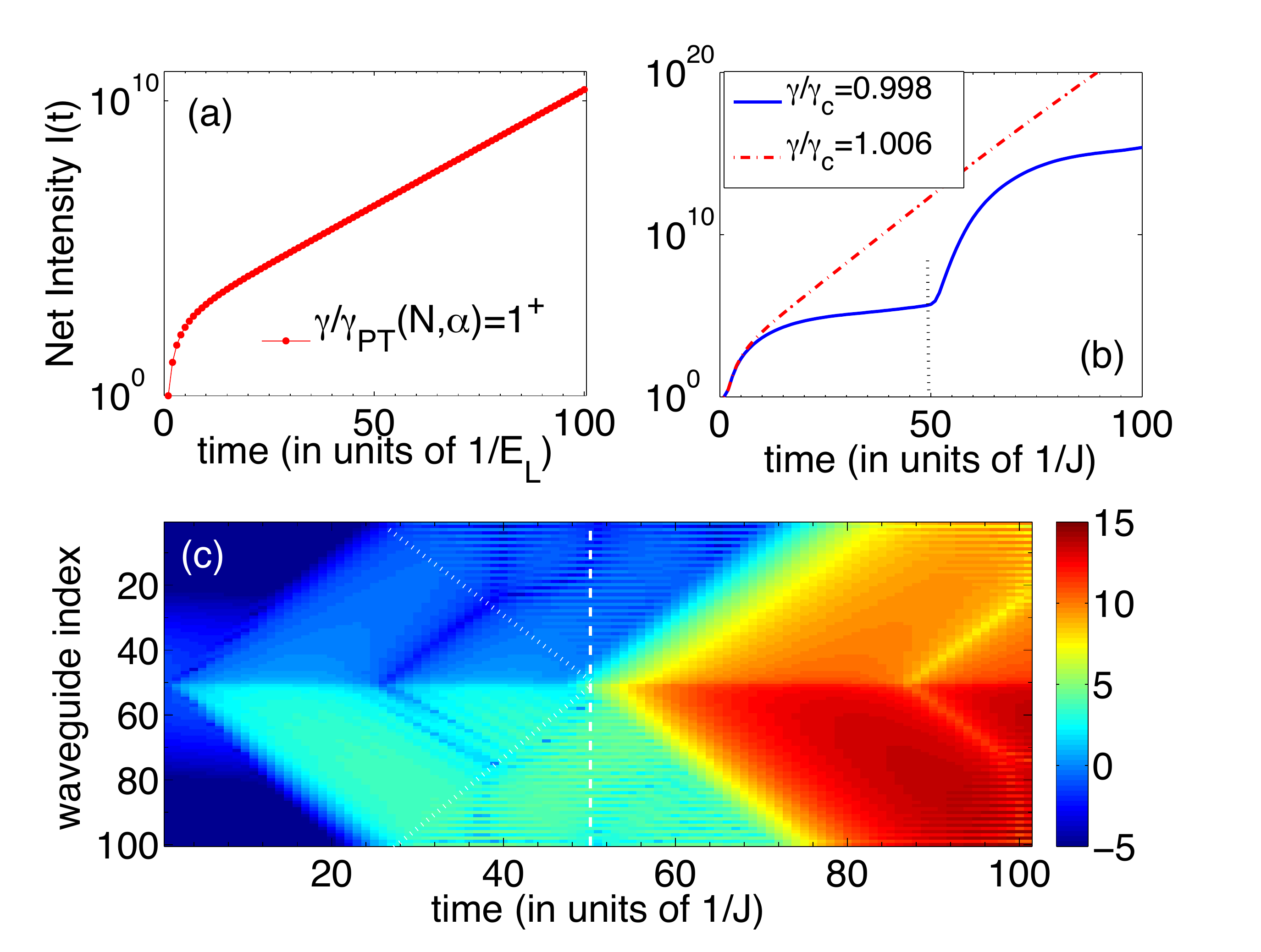}
\caption{(Color online) Panel (a): net intensity $I(t)$ shows exponential growth at long times $tE_L\gg 1$ when the $\mathcal{PT}$-symmetry is broken by extended eigenfunctions. Panel (b): for $\gamma_{PT}(N,\alpha)\ll \gamma<\gamma_c(\alpha)$, the net intensity $I(t)$ shows a staircase structure over time-scale $1/J \ll 1/E_L=(1/J)(N/2)^2$ (blue solid line). This staircase changes to a straight line {\it over the same time-scale} when the bound-state threshold is crossed, $\gamma>\gamma_c(\alpha)$ (red dot-dashed line). Panel (c): $\log I(k,t)$ shows that step in $I(t)$ corresponds to the return of partial waves from the lattice edges at time $\sim N/2J=50/J$ marked by the vertical dashed white line.}
\label{fig:test}
\end{figure}
Lastly, we demonstrate the dramatic effect of the emergence of a localized state with complex energy on the time-scale that determines the ``long-time'' behavior of net intensity $I(t)$ in the $\mathcal{PT}$-symmetry broken phase. All results in Fig.~\ref{fig:test} are obtained for an $N=100$ lattice with $\alpha=-1$, and a broad initial state at the center of the lattice, Eq.(\ref{eq:psi0}), with $\sigma=10$. Panel (a) in Fig.~\ref{fig:test} shows that the net intensity $I(t)$ increases exponentially at times $t E_L\gtrsim 20$ where $1/E_L$ represents the time-scale associated with extended states that break the $\mathcal{PT}$ symmetry. Recall that this time-scale is much longer than the bound-state time-scale, $1/E_L=(1/J)(N/2)^2\gg (1/J)$. Panel (b) in Fig.~\ref{fig:test} shows the behavior of net intensity $I(t)$ over time-scale $1/J$. Below the bound-state threshold, $\gamma/\gamma_{c}=0.998$ (blue solid line),  the intensity on the logarithmic scale shows a step-like structure. This step-structure is replaced by a straight line just above the threshold, $\gamma/\gamma_c=1.006$ (red dot-dashed line). Thus, {\it the presence of a localized state with complex energy dramatically shortens the time-scale for exponential intensity behavior} from $\sim 1/E_L$ to $\sim 1/J\ll 1/E_L$. Panel (c) in Fig.~\ref{fig:test} shows the logarithm of site and time-dependent intensity $I(k,t)$ just below the bound-state threshold, $\gamma/\gamma_c=0.998$. The vertical white dashed line marks the time when the first partial waves, starting near the center of the lattice, return to the center after reflection at the two ends of the open lattice; these return reflections are denoted by white dotted lines. Panel (c) $I(k,t)$ and panel (b) $I(t)$ show that the step-structure in the intensity corresponds to the return of such partial waves. Below the bound-state threshold, such staircase structure in the net intensity $I(t)$ is exhibited at times $T_n=nN/2J$.  


\section{Discussion}
\label{sec:disc}
In this paper, we have investigated $\mathcal{PT}$-symmetry breaking in the presence of extended potentials on a lattice, some of which map onto divergent potentials on a finite segment in the continuum limit. We have shown that the vanishing $\mathcal{PT}$-breaking threshold in lattice models with extended loss-gain potentials guarantees a positive, finite threshold in their continuum counterparts. In addition, we have found that divergent, loss-gain potentials such as $V(x)=i\Gamma L/x$ on a finite segment have a positive $\mathcal{PT}$-breaking threshold. We have shown that the emergence of localized states in $\mathcal{PT}$ potentials dramatically shortens the time-scale necessary for the net intensity to exhibit an exponential-in-time behavior. 

Our results elucidate the connection between lattice and continuum models. They raise similar questions about $\mathcal{PT}$-symmetry breaking in an infinite lattice and its counterpart on an infinite line, as well as a lattice with continuous, local degree of freedom and its field-theory counterpart~\cite{klevansky}. They also hint at the existence of analytical solutions for special values of $\alpha$, such as $\alpha=-2$ or $\alpha=-1$. Addressing these questions will deepen our understanding of $\mathcal{PT}$-symmetry breaking and its observable consequences in experimentally accessible finite lattice systems such as optical waveguide arrays. 

\section*{Acknowledgments}
DS and YJ thank Los Alamos National Laboratory where this work began. This work was supported in part by the National Science Foundation under Grant No. NSF PHY11-25915 and DMR-1054020, and in part by the US Department of Energy. 



\end{document}